\documentstyle[aps,prb,multicol,graphicx]{revtex}
\makeindex
\def\Tr{\mbox{Tr}}
\def\D{\mbox{D}}

\def\max{\mbox{max}}
\def\S{\mbox{S}}

\def\I{\mbox{I}}
\def\M{\mbox{M}}
\def\B{\mbox{B}}

\def\Det{\mbox{Det}}

\title{High cumulants of current fluctuations \\    
out of equilibrium}
\author{D.B. Gutman$^1$, Yuval Gefen$^1$ and A.D. Mirlin$^2$}
\address{$^1$ Department of Condensed Matter Physics,
The Weizmann Institute of Science
Rehovot 76100, Israel \newline
$^2$ Institut f\"ur Nanotechnologie, 
Forschungszentrum Karlsruhe,  
76021 Karlsruhe, 
Germany}
\begin{document}
\date{\today}
\maketitle
\begin{abstract}
We consider  high order current cumulants 
in disordered systems out of equilibrium.
They are interesting and reveal information 
which is not easily exposed by the traditional shot noise.
Despite the fact that the dynamics  of the electrons is classical, 
the standard kinetic theory of fluctuations needs to be modified
to account for those cumulants.   We perform a quantum-mechanical 
calculation using the Keldysh technique and analyze 
its relation to the quasi classical Boltzmann-Langevin scheme.
We also consider the effect of inelastic scattering.
Strong electron-phonon scattering renders 
the current fluctuations Gaussian, completely suppressing 
the $n>2$ cumulants.
Under  strong electron-electron scattering the 
current fluctuations remain non-Gaussian.
\end{abstract}
\begin{multicols}{2}
\section{INTRODUCTION}
\label{introduction}
The fact that electric current exhibits time dependent fluctuations
has been known since the early of $20^{\rm th}$ century.
Still it remains an active field of experimental 
and theoretical research \cite{Blanter}.
Among other features, non-equilibrium shot noise can teach us about 
the rich many-body physics of the  electrons, 
and may serve as a tool to determine 
the effective charge of the elementary carriers. 
While the problem of  non-interacting 
electrons  is practically resolved by now, 
the issue of noise in systems  of interacting electrons remains 
largely an open problem.
In this paper we focus on 
two main issues:
(1) The role of both inelastic electron-electron 
and electron-phonon scattering
for current correlations.
(2) In the absence of interference effects one is tempted to employ 
the semi-classical kinetic theory of fluctuations \cite{Shulman,GGK},
also known as the Boltzmann-Langevin scheme. 
This approach has been originally developed to study 
pair correlation functions, hence is not naturally devised for higher order 
cumulants. A naive application of this approach turns out problematic.
To study high order cumulants, we first need a reliable scheme, which
is why we resort to a microscopic quantum mechanical approach.
The comparison between microscopic calculations  
and the application of the  Boltzmann-Langevin scheme for the study 
of higher order cumulants is the second objective 
of this paper.

The outline of this work is the following:
Section \ref{introduction} addresses a few introductory issues 
concerning current fluctuations in non-interacting systems.
These brief reminders are necessary for the following analysis.
In Section \ref{stoch_model} we consider the two main sources of noise 
in non-interacting systems: thermal fluctuations 
in the contact reservoirs and the stochasticity of the elastic  
scattering process involved.
For this purpose a simple stochastic model is studied.
In  Section \ref{interest} we explain why 
the study of high order correlation functions is 
of interest.
In Section \ref{background} we recall some elements of 
the kinetic theory of fluctuations, explaining the difficulty 
in applying it to high cumulants.
Section \ref{third cumulant} is devoted to the analysis of
the third order current cumulant.
We consider two limiting cases:  that of independent
electrons (\ref{s3_non_interacting}), 
and that of high electron-electron collision rate (\ref{sec:inelastic}).
In Section \ref{comparison} we compare the results 
of the microscopic quantum mechanical analysis with those 
of the Boltzmann-Langevin scheme.
In Section \ref{counting}
we discuss  higher moments of  current cumulants
the full counting statistics, 
and comment briefly on the effect of electron-phonon scattering.
\subsection{Noise In Non-Interacting Systems:
Probabilistic Scattering And Thermal Fluctuations}
\label{stoch_model}
Before discussing the problem of interacting electrons
we would like to recall some important  features 
of fluctuations in a non-interacting electron gas
(for a review see \cite{Blanter}).
One of the issues addressed in the present 
work is  whether what has become to be known as 
``{\it quantum noise} '' can be properly discussed within the (semi)-classical
framework of the kinetic theory of fluctuations.
Surely, non-equilibrium shot noise depends on the
discreteness of the elementary charge carriers
(and this charge is quantized). Also, at low temperatures one
is required to employ the Fermi-Dirac statistics  governing the occupation of the reservoir states.
In addition, the channel transmission and reflection probabilities are governed by quantum mechanics.
But other than that, interference effects appear to play a very minor
role in the formation of current fluctuations.
Hence the appeal of a semi-classical approach. 
To  understand the origin of the current fluctuations 
we consider a simple stochastic model, following earlier works  
\cite{Levitov_Lesovik,Levitov_Reznikov,Reznikov_private}.
This model is a caricature of the physics underlying 
low frequency transport through a quantum point contact 
(QPC) with a single channel having a transmission probability ${\cal{T}}$. 
The QPC is connected to two reservoirs 
with respective chemical potentials $\mu_L$ and $\mu_R$ 
\mbox{(this model is readily generalized to the multi-channel case)}.
\end{multicols}   
\begin{figure}[h]
\includegraphics[width=0.7\textwidth]{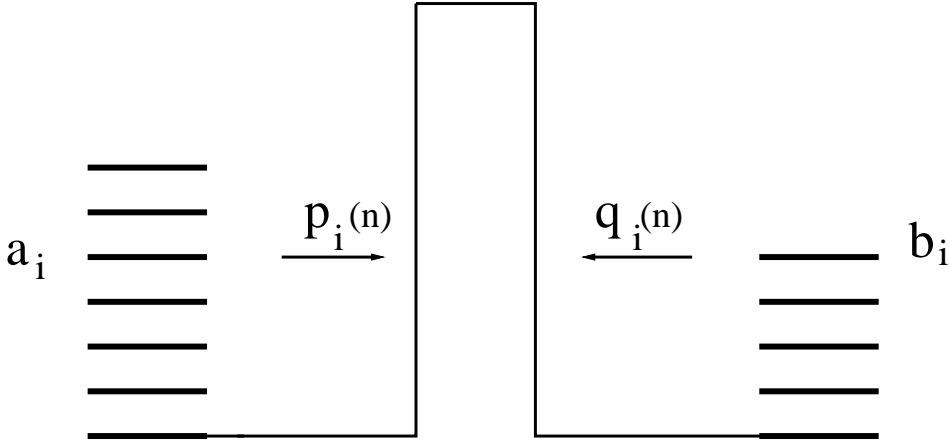}
\caption{A simplified view of transport through a single
channel quantum point contact. 
For definitions consult the text.}
\label{fig2}
\end{figure}
\begin{multicols}{2}
For our needs electrons occupying (in principle broadened) 
single particle levels (depicted in a Fig.\ref{fig2})
can  be perceived as  classical particles 
attempting to pass through the barrier
(with transmission probability ${\cal{T}}$).
In principle ${\cal{T}}$ may be a function of the level's energy 
(${\cal{T}} \rightarrow {\cal{T}}_i$).
For the mean level spacing $\Delta \epsilon$ the time 
$\Delta t=h/\Delta \epsilon$ can be interpreted as
a time  interval between two consecutive attempts.
Since the electron charge is discrete  
(and is well-localized on either the left or the right side of the barrier)
the  results of a transmission attempts have a binomial statistics.

Consider the fluctuation in the net charge transmitted through
the QPC over the time interval $\bar{t}\gg \Delta t$. 
The number of attempts $N=\bar{t}/\Delta t$ occurring within this interval 
is large.
Since the attempts are discrete events, they can be enumerated.  
To  describe the result of the $n^{\rm th}$ attempt  
of an  electron originating from level $i$ on the  left (right) 
we define  the quantity 
$p_i(n)$ and $q_i(n)$ (cf. Fig.\ref{fig2}).
For an attempt ending up in the transmission of an 
electron from left to right  (from right to left)  
the quantity $p_i(n)$  ($q_i(n)$) is  assigned the  value $1$. 
Otherwise $p_i(n)$ ($q_i(n)$)  are zero. 
This implies that $\{p_i(n)\}$'s and $\{q_i(n)\}$'s have each 
binomial statistics
with the  expectation value 
$\langle p_i(n)\rangle=\langle q_i(n)\rangle={\cal{T}}_i$. 
We also assume that neither different attempts from the same level,
nor attempts of the electrons coming 
from different levels are correlated.
In addition to fluctuations in the transmission process,
there are also  fluctuations of the occupation numbers $a_i$ 
($b_i$) of a single particle level $i$. 
These represent thermal fluctuations in the reservoirs. 
For fermions  they possess a binomial distribution \cite{LL}.
We end up with 
\begin{eqnarray}&&
P(x)=(1-\langle x\rangle)\delta(x)+\langle x\rangle\delta(1-x) \nonumber \\&&
{\rm for} \ \ x=p_i(n), q_i(n), a_i, b_i,
\end{eqnarray}
and the correlations
\begin{eqnarray}&&
\langle p_i(n)p_j(n')\rangle=\delta(i,j)\delta(n,n'){\cal{T}}_i; 
\nonumber \\&&
\langle q_i(n)q_j(n')\rangle=\delta(i,j)\delta(n,n'){\cal{T}}_i. 
\label{pp}
\end{eqnarray}
Also
\begin{eqnarray}&&
\langle a_i a_j\rangle=\left\{
\begin{array}{l}
n^L_i, \,\,\,\, i=j 
\\
0, \,\,\,\,\,\,\,\, i\neq j.
\end{array}
\right.
\,\,\,\,
\langle b_i b_j\rangle=\left\{
\begin{array}{l}
n^R_i, \,\,\,\, i=j 
\\
0, \,\,\,\,\,\,\,\, i\neq j.
\end{array}
\right.
\label{aa}
\end{eqnarray}
Here $n_i^L\equiv\langle a_i\rangle$ and $n_i^R\equiv\langle b_i\rangle$.
The typical time scale over which the occupation number of a level
fluctuates 
is usually dictated by the interaction among the electrons
or with external agents.
We assume that at the leads the  time fluctuations of the occupation 
numbers ($a_i$ and $b_i$)
over the interval $\bar{t}$ are negligible.

Motivated by this picture, we consider 
fluctuations in the net number $Q_{\bar{t}}$  of electrons  
transmitted through the constriction within the time interval $\bar{t}$.
Employing the Pauli exclusion principle one can write 
\begin{eqnarray}
\label{st_model1}
Q_{\bar{t}}=
\sum_{n=1}^N\bigg[\sum_i [p_i(n)a_i(1-b_i)-q_i(n)b_i(1-a_i)]\bigg].
\end{eqnarray}

According to eqs.(\ref{pp},\ref{aa} and \ref{st_model1}) the expectation 
value of the transmitted charge is
\begin{eqnarray}&&
\label{st_model4}
\langle{Q}_{\bar{t}}\rangle={\cal{T}}N\sum_i [n^L_i(1-n^R_i)-n^R_i(1-n^L_i)].
\end{eqnarray}
Evidently it is  related to the value of the  d.c. current through 
$\langle Q_{\bar{t}}\rangle =\bar{t}I/e$.
We next discuss the higher order moments of
current fluctuations.
As was shown in Refs.\cite{Quantum_Galvanometr} (see also \cite{Klich})
the experimentally measured high order current cumulants
should be defined in quite a subtle manner.
To make contact between the measured observables and theoretically calculated quantities, we recall that within the  Keldysh formalism \cite{Keldysh} 
the time axes is folded. For any
given moment of time there are two current operators,
one for the upper and another for the lower 
branch of the contour.
The product of the symmetric combination of these two operators, $I_2$, 
time ordered along the Keldysh contour ($T_c$) and averaged
with respect to the density matrix yields the proper
current correlation function. 
In a stationary situation the pair current correlation
function \cite{Gavish}  
\begin{eqnarray}&&
\S_2(t-t')=\langle T_c I_2(t)I_2(t')\rangle
\end{eqnarray}
depends only on the difference between the time indices.
Similarly we define the third order current correlation function 
\begin{eqnarray}&&
\S_3(t-t';t'-t'')=\langle T_c I_2 (t)I_2 (t')I(t'')\rangle \, .
\end{eqnarray}
In  Fourier space it can be represented as
\begin{eqnarray}&&
\hspace{-1.5cm}\S_2(\omega_1)=\int_c d(t-t') 
e^{-i\omega_1(t-t')}\S_2(t-t')\,\, , \\&&
\hspace{-1.5cm}\S_3(\omega_1,\omega_2)\!\!=\!\!\int_c\!d(\!t\!-\!t'\!)\!d(\!t'\!-\!t''\!)\!e^{-i\omega_1\!(t\!-t'\!)\!-\!i\omega_2(\!t'-t''\!)}\!\S_3(\!t-\!t'\!,t'-t'').\nonumber
\end{eqnarray}
Usually it is more interesting to consider 
only the cumulants, i.e. 
irreducible parts of the correlation functions.
For the pair current correlation function we have
\begin{eqnarray}&&
{\cal{S}}_2(\omega=0)=\S_2-I^2 , 
\end{eqnarray}
while the third order current cumulant is given by 
\begin{eqnarray}&&
\label{z38}
{\cal{S}}_3(\omega_1=0,\omega_2=0)=\S_3-3I\S_2+2I^3,
\end{eqnarray}
where $\S_2,\S_3$ are taken at zero frequencies.
Microscopic calculations of the current cumulants will 
be performed in the Sections \ref{third cumulant}  and \ref{counting} 
below.
Here we evaluate the 
cumulants within the stochastic model described above.
The results agree with the low frequency
current cumulants of the non-interacting electrons 
in the  QPC \cite{Levitov_Lesovik}.
The temperature dependence of those cumulants is
qualitatively similar to the dependence of the
current cumulants in a disordered junction,
analyzed below.
 
To find the variance  of the stochastic variable $Q$ 
we employ  eqs.(\ref{pp}, \ref{aa} and \ref{st_model1}) 
\begin{eqnarray}&&
\langle Q^2_{\bar{t}}\rangle-\langle Q_{\bar{t}}\rangle^2=
{\cal{T}}(1-{\cal{T}})\sum_i[n_L^2(i)-n_R^2(i)]^2+\nonumber \\&&
{\cal{T}}\sum_i[n_L(i)+n_R(i)-n_L^2(i)-n_R^2(i)]
 .
\label{st_model6}
\end{eqnarray}
The first term on the r.h.s. of eq.(\ref{st_model6}) vanishes at 
thermal equilibrium. 
We associate this part with the shot noise of the electrons.
The second term vanishes at zero temperature, 
and we associate it with the thermal noise.
As expected the ``shot noise'' part vanishes in the limit of perfect conductor (${\cal{T}} \rightarrow 1$).
Going along the same procedure 
for the third order cumulant one obtains \cite{Levitov_Reznikov}
in the limit of low temperature ($eV \gg T$)
\begin{eqnarray}&&
\langle (Q_{\bar{t}}-\langle Q_{\bar{t}}\rangle)^3\rangle={\cal{T}}(1-{\cal{T}})(1-2{\cal{T}})\langle Q_{\bar{t}}\rangle
\label{st_model7}
\end{eqnarray}
and in the limit of high temperature ($eV \ll T$)
\begin{eqnarray}&&
\label{st_model8}
\langle (Q_{\bar{t}}- \langle Q_{\bar{t}}\rangle)^3 \rangle={\cal{T}}(1-{\cal{T}})\langle Q_{\bar{t}}\rangle .
\end{eqnarray}
Note that with increasing the temperature the third order cumulant of
the transmitted charge approaches a constant.
This is  a robust feature of all odd order cumulants,
which can be understood from quite general arguments.
Indeed, since the current operator changes sign under time reversal
transformation, any even-order correlation function of the current
fluctuations (e.g. ${\cal{S}}_2$) taken at zero frequency is
invariant under this operation. Assuming that current correlators
are functions of the average current, $I$, it follows that
even-order   correlation functions depend only on the absolute
value of the electric current (and are independent of the
direction of the current). In the Ohmic regime this  means that
even-order current correlation functions (at zero frequency)  are
even functions of the applied voltage. Evidently, this general
observation  agrees with the result eq.~(\ref{st_model6}).

By contrast,  odd-order current correlation functions change their
sign under time reversal transformation. In other words, such
correlation functions depend on the direction of the current, and
not only on its absolute value. Therefore in  the Ohmic regime,
odd-order correlation functions of current are odd-order functions
of the applied voltage. This condition automatically guarantees
that odd-order correlation functions {\it vanish}  at thermal
equilibrium.

One can show that by considering high order moments of the
stochastic model one  reproduces the correct 
results for any cumulants of a current noise in a QPC.
Of course, the solution of a toy  model can not replace the
real microscopic calculations. 
But the fact that the results of the
latter and the stochastic model agree 
suggests that the underlying physics 
is rather simple (and it is basically captured by such a simple model).
It is the combination of thermal fluctuations 
(in the occupation of the single electron states) and random transmission 
of particles through the barrier that gives rise to current fluctuations.
These two sources of stochasticity remain there when the electron 
can no longer be considered  non-interacting. In that case, though,
one cannot consider fluctuations at different 
energy levels to be independent.
\subsection{Why Are High Order Cumulants Interesting?}
\label{interest}
Low frequency current fluctuations give rise 
to a large (in general infinite) number of 
irreducible correlation functions (cumulants).
The pair current  correlation function 
provides us with only partial information about 
current fluctuations. To obtain the complete 
picture one should consider  high order cumulants as well.
As we have explained in Section \ref{stoch_model}
the symmetry-dictated properties 
of odd and even correlation function are 
very different from each other; in particular odd order current cumulants  are
not masked by thermal fluctuations.
For this reason  they can be used 
for probing non-equilibrium properties at relatively high 
temperatures. 
Shot noise has been used to detect an effective quasi-particle charges 
in the FQHE regime \cite{Reznikov,Glattli}.
Potentially, odd moments can be used to measure the effective 
quasiparticle charge in other 
strongly-correlated systems \cite{Reznikov_private}. 
This may be needed for systems undergoing a 
transition controlled by  temperature  
(for example normal-super-conductor)
and having different fundamental excitations 
at  different temperature regimes.
One may hope therefore, that better understanding
of current correlations will teach us more 
about the many-body electron physics.
At this moment this remains a challenge.
Before trying to reach this goal, 
we need to understand the genuine properties
of the high order cumulant in the relatively simple
physical models. This is done next.
\subsection{Background and Issues to Be Discussed}
\label{background}
As was mentioned above it is quite
appealing to try to discuss current noise in terms 
of the semi-classical kinetic equation.
To be more specific, let us consider a disordered metallic constriction. 
Its length $L$ is much greater
than the elastic mean free path $l$.
The disorder inside the constriction is short-ranged, 
weak and uncorrelated.
Under these conditions the 
electrons kinetics can be described by the Boltzmann equation:

\begin{eqnarray}&&
\label{int_eq7}
\hat{{\cal{L}}} \bar{f}({\bf p}, {\bf r},t)={\rm Col\{\bar{f}\}}. 
\end{eqnarray}
Here ${\rm Col\{\bar{f}\}}$ is the collision integral and
\begin{eqnarray}&&
\label{int_eq88}
\hat{{\cal{L}}}=\left(\frac{\partial}{\partial t}+{\bf v}\cdot\nabla_{\bf r}+
e{\bf E}\cdot\nabla_{\bf p}\right)
\end{eqnarray}
is the Liouville  operator of a particle moving in 
an external electric field ${\bf E}$.
The pair correlation function of any macroscopic quantity 
may be found from the kinetic theory of fluctuations.
Within this theory the distribution function ($f$) consists
of the coarse-grained ($\bar{f}$) and fluctuating ( $\delta f$) parts
\begin{eqnarray}&&
\label{int_eq89}
f({\bf p}, {\bf r},t)=\bar{f}({\bf p}, {\bf r},t)+\delta f({\bf p}, {\bf r},t).
\end{eqnarray}
Since $f$ is a macroscopic quantity (a quantity 
associated with a large number of particles) 
it must satisfy the Onsager's regression hypothesis 
(see Ref. \cite{LP}, Section 19).
 
To sketch this hypothesis for the case of interest
we consider a  perturbation of the equilibrium distribution function 
$\delta \bar{f}({\bf p,r},t)$, small but substantially larger 
than the typical fluctuations of the distribution function.
It follows then that the distribution function 
(with a high probability)
will  evolve toward the equilibrium state.
Its relaxation dynamics is governed by the Boltzmann equation 
and  because the perturbation is small,
the collision integral can  be linearized.
According to Onsager the correlation function  of  {\it any}
macroscopic quantity, and in particular $\langle \delta f({\bf p,r},t) f(\bf{p',r'},t')\rangle$,
is governed by the {\it same} equation as the one governing  its relaxation  
(i.e. as equation governing the quantity $\delta \bar{f}({\bf p,r},t)$)
\begin{eqnarray}&&
\label{int_eq6}
\left(\hat{{\cal{L}}}({\bf p}, {\bf r},t)+{\cal{I}}({\bf p}, {\bf r},t)
\right) \langle\delta f({\bf p}, {\bf r},t)\delta f({\bf p'}, {\bf r'},t')\rangle =0, \nonumber \\&&
{\rm for }  \ \ \ \ \ \ t>t'.
\end{eqnarray}
Here ${\cal{I}}$ is a linearized collision integral. 
It was later suggested by Lax \cite{Lax} 
(and can be proven within Keldysh formalism) 
that  eq.(\ref{int_eq6}) does hold for any  
stationary, not necessarily equilibrium, state.
However, we need to recall that the Onsager hypothesis was formulated
only for a {\it pair} correlation function.
There is no obvious way to apply this logic for higher cumulants. 

An alternative route of describing fluctuations
within kinetic theory which is seemingly free of this difficulty
had been proposed by Kogan and Shul'man\cite{Shulman}.
Their picture is the following.
The real space is divided into small
volumes (as explained in a Ref.\cite{LP}). 
The function $\bar{f}({\bf p}, {\bf r},t)$ represents the average 
number of particles in the state ${\bf p}$ of a unit volume element (cell)
labeled by index (${\bf r}$).
The total number of the electron in every cell must be large. 
It was suggested in Ref. \cite{Shulman} that fluctuations
of this number can be taken into account by adding a random (Langevin) 
source term to the Boltzmann equation.
The resulting stochastic equation (including this {\it additive} noise)
is called the Boltzmann-Langevin equation.
\begin{eqnarray}
\label{Boltzmann_Langevin}
\left(\hat{{\cal{L}}}({\bf p}, {\bf r},t)+{\cal{I}}({\bf p}, {\bf r},t)
\right) \delta f({\bf p}, {\bf r},t)=\delta J({\bf p}, {\bf r}, t)\, .
\end{eqnarray}
The Langevin source $\delta J({\bf p}, {\bf r},t)$ 
denotes a random number of particles incoming  
into the given state in some interval (around time $t$).  
Since eq.(\ref{Boltzmann_Langevin}) is a linear one, 
the  statistics of the  distribution function is determined 
by the  random source term $\delta J$.
To establish its properties Kogan and Shul'man had
used a rather simple physical picture.
To be consistent with the Boltzmann equation they have assumed 
that interference effects are weak.
The collision events are local in space and time.
Since the typical number of electrons inside the cell is large
one can  ignore the correlation  between the scattering
of different electrons. 
The electron scattering
is a Poissonian process, with the number of scattered 
particles  within any given cell
(over a microscopic time interval) being large.
While this picture yields a correct result for 
the pair correlation function, it substantially 
underestimates all high order
correlators (starting from ${\cal{S}}_3$). 

\section{THIRD ORDER CURRENT CUMULANT}
\label{third cumulant}
In this section we use microscopical calculations to evaluate 
the third order current cumulant for a quasi one-dimensional system
of a length $L$  with diffusive disorder\cite{GG2}.
We start with the coordinate-dependent correlation function 
\begin{eqnarray}
\label{s3}
\S_3(x,t;x',t';x'',t'')=\langle T_c \hat{I}_2(x,t)\hat{I}_2(x',t')\hat{I}_2(x'',t'') \rangle \,\, .
\end{eqnarray}
Here $x$ is a coordinate measured along a quasi one-dimensional
wire ($0 \le x\le L$) of cross-section ${\cal{A}}$;  $T_c$ is the
time ordering operator along the Keldysh contour.
Next we perform  the Fourier transform 
with respect to the time difference as in eq.(\ref{s3_fourier}).
For small values of the  frequencies $\omega_1,\omega_2$,
(small compared with the inverse diffusion time along the wire),
the current fluctuations are independent of the spatial
coordinate.  We next evaluate the expression, eq.~(\ref{s3}), for
a disordered junction, in the hope that the qualitative properties
we are after are  not strongly system dependent. In the present
section we consider non-interacting electrons in the presence of a
short-range, delta correlated and weak disorder potential
($\epsilon_f \tau \gg \hbar$, where $\tau$ is the elastic mean
free time and $\epsilon_f$ is the  Fermi energy). To calculate
$\S_3$ we employ the $\sigma$-model formalism, recently put
forward for dealing with non-equilibrium diffusive systems (for
details see Ref.\cite{GG}). 

The disorder potential is
$\delta$-correlated:
\begin{equation}
\label{<>}
\langle U_{\rm dis}({\bf r})~U_{\rm dis}({\bf r}')\rangle = \frac{1}{2\pi \nu \tau} ~
\delta({\bf r} - {\bf r}'),
\end{equation}
where $\nu$ is the  density of states at the Fermi energy.

The Hamiltonian we are concerned with is:
\begin{eqnarray}&&
H=H_0+H_{\rm int}.
\end{eqnarray}
The motion of electrons in the disorder potential is described by:
\begin{eqnarray}
\label{Hamiltonian} H_0=\int_{\rm Volume} d{\bf r} \bar{\Psi}({\bf
r})\bigg[-\frac{\hbar^2}{2m}(\nabla-i{\bf a})^2 + U_{\rm
dis}\bigg]\Psi({\bf r}).
\end{eqnarray}
Here $c{\bf a}/e$ is a vector potential.
The Coulomb interaction among the electrons is described by
\begin{eqnarray}&&
H_{\rm int}=\frac{1}{2} \int d{\bf r} d{\bf r'} \bar\Psi({\bf r})
\bar\Psi({\bf r}') V_0({\bf r}-{\bf r}') \Psi({\bf r})\Psi({\bf
r}')\,\, ,
\end{eqnarray}
where
\begin{eqnarray}&&
V_0({\bf r}-{\bf r}')=\frac{e^2}{|{\bf r}-{\bf r}'|} \,\,.
\end{eqnarray}

\subsection{Weak Inelastic Collisions }
\label{s3_non_interacting}
Following the procedure outlined in Ref. \cite{GG}, we 
introduce a generating functional and average it over disorder.
Next we perform a Hubbard-Stratonovich transformation,
integrating out fermionic degrees of freedom.
Employing the diffusive approximation 
one obtains an effective generating functional
expressed  as a path integral over a
bosonic matrix field $Q$
\begin{eqnarray}&&
\label{gen1} Z[{\bf a}] =\int {\cal D}Q \exp(iS[Q,{\bf a}])\,\, .
\end{eqnarray}
Here the integration is performed  over the manifold
\begin{eqnarray}&&
\label{manifold}
\int Q(x,t,t_1)Q(x,t_1,t')dt_1=\delta(t-t'),
\end{eqnarray}
the effective action is given by
\begin{eqnarray}&&
\label{action1} iS[Q,{\bf a}]=-\frac{\pi\hbar\nu}{4}\Tr\big\{D \left(
\nabla Q+i[{\bf a}_\alpha \gamma^\alpha,Q] \right)^2 +4i
\hat{\epsilon} Q\big\}\,\,,
\end{eqnarray}

and
\begin{eqnarray}&&
\gamma_1=\left(\matrix {1 & 0\cr 0 & 1 \cr}\right) , \;
\gamma_2=\left(\matrix {0 & 1\cr 1 & 0 \cr}\right) \, .
\end{eqnarray}
$\Tr$ represents summation over all spatio-temporal and Keldysh components.
Here ${\bf a}_1$ and ${\bf a}_2$ are the Keldysh rotated classical  and quantum components of ${\bf a}$. Hereafter we focus our attention on ${\bf a}_1$, ${\bf a}_2$, the components in the direction along the wire.
The third order current correlator may now be expressed as
functional differentiation of the generating functional $Z[a]$
with respect to $a_2$
\begin{eqnarray}
\label{z27}
\!\S_3(\!t_1\!-\!t_2,t_2\!-\!t_3\!)\!=\frac{ie^3}{8}\frac{\delta^3Z[a]}{\delta\!a_2(x_1,t_1)\delta\!a_2(x_2,t_2)\delta\!a_2(x_3,t_3)}.
\end{eqnarray}
Performing this  functional differentiation  one obtains the
following result
\begin{eqnarray}&&
\label{z28} 
\S_3(\!t_1\!-\!t_2,t_2\!-\!t_3\!)\!
=\frac{e^3{\cal{A}}(\pi\hbar\nu D)^2}{16}
\bigg\langle\hat{\M}(x_1,t_1)\hat{\I}^D
(x_2,x_3,t_2,t_3)+\nonumber \\&&
(x_1\!,\!t_1\!\leftrightarrow\!x_3,t_3)\!+\!
(x_1\!,\!t_1\!\leftrightarrow\!x_2,t_2)\!+\nonumber \\&&
\frac{\pi\hbar\nu\!D}{4}\hat{\M}(x_1,t_1)\hat{\M}(x_2,\!t_2)\hat{\M}(\!x_3,t_3)
\bigg\rangle .
\end{eqnarray}
Here we have defined
\begin{eqnarray}&&
\label{e33}
\hat{I}^D(x,x',t,t')=\Tr^K\!\bigg\{\!Q_{x,t,t'}\gamma_2Q_{x',t',t}\gamma_2-
\delta_{t,t'}\gamma_1\bigg\}\delta_{x,x'} ,
\end{eqnarray}
\begin{eqnarray}&&
\label{M}
\hat{\M}(x,t)=\Tr^K\!\bigg\{\!\int\!dt_1\bigg([Q_{x,t,t_1};\nabla ]~ Q_{x,t_1,t}
\bigg)\gamma_2\bigg\}.
\end{eqnarray}
We employ the notations $Q(x,t,t')\equiv Q_{x,t,t'}$; $\Tr^K$ is
the trace  taken with respect to the Keldysh indices; $\langle
\rangle$ denotes a quantum-mechanical  expectation value. The
matrix $Q$ can be parameterized as
\begin{eqnarray}&&
\label{e24} Q=\Lambda\exp\left(W\right) \,\, ,
{\rm where} \ \, \ 
\Lambda W+W \Lambda=0 
\end{eqnarray}
and $\Lambda$ is the saddle point of the action (\ref{action1})
\begin{eqnarray}&&
\label{e25} \Lambda(x,\epsilon)=\left(\matrix{1 & 2F(x,\epsilon)
\cr 0 & -1\cr}\right)\,\, .
\end{eqnarray}
The function $F$ is related to the single particle distribution
function $f$ through
\begin{eqnarray}&&
\label{e31}
F(x,\epsilon)=1-2f(x,\epsilon) \,\, .
\end{eqnarray}
The matrix $W_{x,\epsilon,\epsilon'}$,  in turn, is parameterized
as follows:
\begin{eqnarray}
W_{x,\epsilon,\epsilon'}\!=
\!\left(
\matrix {F_{x,\epsilon}
\bar{w}_{x,\epsilon,\epsilon'} & -w_{x,\epsilon,\epsilon'} + F_{x,\epsilon}\bar{w}_{x,\epsilon,\epsilon'}F_{x,\epsilon'}\cr-\bar{w}_{x,\epsilon,\epsilon'}  & - \bar{w}_{x,\epsilon,\epsilon'}F_{x,\epsilon'}\cr}
\right)  .
\end{eqnarray}
It is convenient to introduce the diffusion propagator
\begin{eqnarray}&&
(-i\omega +D\nabla^2)\D(x,x'\omega)=\frac{1}{\pi\hbar\nu}\delta(x-x')
\, \, .
\end{eqnarray}
The absence of diffusive motion   in  clean metallic leads
implies that the  diffusion propagator must vanish at the end
points of the constriction. In addition,  there is  no current
flowing in the transversal direction (hard wall boundary
conditions). It follows that the component of the gradient of the
diffusion propagator in that direction (calculated at the hard
wall edges) must vanish as well. The correlation functions of the
fields $w, \bar{w}$ are then given by:
\begin{eqnarray}&&
\label{corr1}
\langle w(x,\epsilon_1,\epsilon_2) \bar{w}(x',\epsilon_3,\epsilon_4)
\rangle=\nonumber \\&&
2(2\pi)^2\delta(\epsilon_1-\epsilon_4)
\delta(\epsilon_2-\epsilon_3) \D(x,x',\epsilon_1-\epsilon_2)
\nonumber \, ,\\&&
\langle w(x,\epsilon_1,\epsilon_2)w(x',\epsilon_3,\epsilon_4)\rangle=-g(2\pi)^3\delta(\epsilon_1-\epsilon_4)
\delta(\epsilon_2-\epsilon_3) \nonumber \\&&
\int dx_1 \D_{\epsilon_1-\epsilon_2,x,x_1}\nabla F_{\epsilon_2,x_1}
\nabla F_{\epsilon_1,x_1} \D_{\epsilon_2-\epsilon_1,x_1,x'} \,\, ,\nonumber \\&&
\langle \bar{w}(x,\epsilon_1,\epsilon_2) \bar{w}(x',\epsilon_3,\epsilon_4)
\rangle=0 \,\, .
\end{eqnarray}

To evaluate  ${\cal{S}}_3$ one follows steps similar to those that led to
the derivation of $\S_2$, see Ref. \cite{GG}. If all relevant
energy scales  in the problem are smaller than the transversal
Thouless energy ($E_{Th}=D/L_T^2$, where $L_T$ is a width of a
wire), the wire is effectively quasi-one dimensional. In that case
only the lowest transversal mode of the diffusive propagator can
be taken into account,
 which yields
\begin{eqnarray}&&
\label{elastic_diffusion}
\D(x_1,x_2)=\frac{1}{2\pi g}
\bigg[|x_1-x_2|-x_1-x_2+\frac{2x_1x_2}{L}\bigg]\, \, .
\end{eqnarray}
Here $g=\hbar\nu D$.
The electron distribution function in this system is equal to
\begin{eqnarray}&&
\label{distribution_function}
F(x,\epsilon)=\frac{x}{L}F_{eq}\left(\epsilon-\frac{eV}{2}\right)+
\left(1-\frac{x}{L}\right)F_{eq}\left(\epsilon+\frac{eV}{2}\right)\,\,.
\end{eqnarray}
The quantities $F$ and $\D$ determine the correlation functions,
eq.~(\ref{corr1}). We can now begin to evaluate $\S_3$, (c.f.
eq.~(\ref{z28})), performing a perturbative expansion in the
fluctuations around the saddle point solution, eq.~(\ref{e25}).
After some algebra we find that in the zero frequency limit the
third order {\it cumulant}  is given by
\begin{eqnarray}&&
{\cal{S}}_3(\omega_1=0,\omega_2=0)=\frac{3e^3{\cal{A}}\pi g^2}{\hbar L^3}
\int_0^L d x_1 dx_2 \nonumber \\&&
\int_{-\infty}^{\infty}d\epsilon F(\epsilon,x_1)
\D[0,x_1,x_2]
\nabla\left(F^2(\epsilon,x_2)\right)
\,\, .
\label{a10}
\end{eqnarray}
Integrating over energies and coordinates we obtain
\begin{eqnarray}&&
\label{e27}
{\cal{S}}_3(\omega_1=0,\omega_2=0)=e^2 I y(p)\,\, , \nonumber \\&&
y(p)=\frac{6(-1+e^{4p})+(1-26e^{2p}+e^{4p})p}{15p(-1+e^{2p})^2} \,\, ,
\end{eqnarray}
where $p=eV/2T$. The function $y$ is depicted in Fig.1,
where it is plotted on a logarithmic scale.
\begin{figure}[h]
\includegraphics[width=0.5\textwidth]{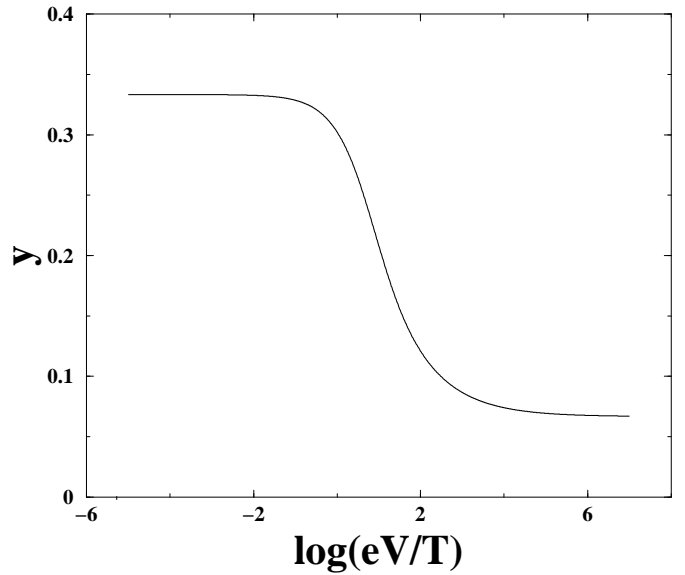}
\label{fig1}
\caption{The scaling function $y$ plotted on a logarithmic scale,
cf. eq.~(\ref{e27})}
\end{figure}

Let us now discuss the main features of the function ${\cal{S}}_3$.
In agreement with symmetry requirements ${\cal{S}}_3$ is an odd
function of the voltage (the even correlator ${\cal{S}}_2$ is
proportional to the absolute value of voltage), and vanishes at
equilibrium. The zero temperature result (high voltage limit) has
already been obtained by means of the scattering states approach
for single-channel systems \cite{Levitov_Lesovik}, and later
generalized by means of Random Matrix Theory (RMT) to
multi-channel systems (chaotic and diffusive) \cite{Levitov_Lee}.
In our derivation we do not assume the applicability of RMT \cite{Beenakker}.
Our result covers the whole temperature range. We obtain that at low
temperatures the third order cumulant  is linear in the voltage
\begin{eqnarray}&&
\label{a11} {\cal{S}}_3=\frac{e^2}{15} I\,\, .
\end{eqnarray}
At high temperatures the electrons in the reservoirs are not
anymore in the ground state, so the correlations are partially
washed out by thermal fluctuations. One may then expand $y(p)$,
eq.~(\ref{e27}), in a  series of $eV/2T$. The leading term in this
high temperature expansion  is linear in the voltage
\begin{eqnarray}&&
\label{a12} {\cal{S}}_3=\frac{e^2}{3} I\,\, .
\end{eqnarray}
Note that although thermal fluctuations enhance the noise
(compared with the zero temperature limit), eqs.(\ref{a11}) and
(\ref{a12}) differ only by a numerical factor. 
The experimental study of ${\cal{S}}_3$ (and higher odd cumulants)  
provides one with a direct
probe of non-equilibrium behavior, not masked by equilibrium
thermal fluctuations.
\subsection{Strong Inelastic Collisions}
\label{sec:inelastic}
In our analysis so far we have completely
ignored  inelastic collisions among the electrons. This
procedure is well justified  provided that the inelastic length
greatly exceeds the system's  size. However, if this is not the case,
different analysis is called for.  To understand why inelastic
collisions do matter for  current fluctuations, we would like
to recall the analysis of ${\cal{S}}_2$ for a similar problem.
 The latter function is fully determined by the effective electron
temperature. Collisions among  electrons, which are subject to an external
bias,  increase the temperature of those electrons. This, in turn,
leads to the  enhancement of ${\cal{S}}_2$, cf. Refs.
\cite{Nagaev95,Kozub}. 
In the limit of short inelastic
length
\begin{equation}
\label{inelastic} l_{\rm e-e} \ll L  \,\, ,
\end{equation}
the zero frequency and zero temperature noise is

\begin{equation}
{\cal{S}}_2(0)=\frac{\sqrt{3}}{4}eI \,\, .
\end{equation}
In the present section we consider the effect of inelastic
electron collisions on ${\cal{S}}_3$.
We  assume that the electron-phonon collision length 
is large, $l_{\rm e-ph} \gg L$,
hence electron-phonon scattering may be neglected.
The Hamiltonian we are
concerned with is
\begin{eqnarray}&&
H=H_0+H_{\rm int}.
\end{eqnarray}
The Coulomb interaction among the electrons is described by
\begin{eqnarray}&&
H_{\rm int}=\frac{1}{2} \int d{\bf r} d{\bf r'} \bar\Psi({\bf r})
\bar\Psi({\bf r}') V_0({\bf r}-{\bf r}') \Psi({\bf r})\Psi({\bf
r}')\,\, ,
\end{eqnarray}
where
\begin{eqnarray}&&
V_0({\bf r}-{\bf r}')=\frac{e^2}{|{\bf r}-{\bf r}'|} \,\,.
\end{eqnarray}
 We need to deal with the  effect of electron-electron
interactions  in the presence of disorder and away from
equilibrium. Following Ref. \cite{KA} one may introduce an
auxiliary bosonic field
\begin{eqnarray}&&
\Phi=\left(
\begin{array}{l}
\phi_1 \\
\phi_2
\end{array} \right),
\end{eqnarray}
which  decouples the interaction in the particle-hole channel.
Now  the partition function (eq.  (\ref{gen1})) is a functional integral
over both  the  bosonic fields $Q$ and $\phi$\, ,
\begin{eqnarray}&&
\label{gen2}
\langle Z \rangle =\int_{Q^2=1} {\cal D}Q {\cal D}\phi\exp(iS_{total})\, \, .
\end{eqnarray}
The action is
\begin{eqnarray}&&
iS_{total}=iS[\Phi]+iS[\Phi,Q] \, , \label{act1} \\&&
iS[\Phi]=i\Tr\{\Phi ^T V^{-1}_0\gamma^2\Phi\}  \, , \label{act2} \\&&
iS[\Phi,Q]=-\frac{\pi\nu}{4\tau}\Tr\{Q^2\}\!+
\Tr\ln \Big[\hat G_0^{-1}\!+\frac{iQ}{2\tau}+\phi_{\alpha}\gamma^{\alpha}
\Big].  \label{act3}
\end{eqnarray}
It is convenient to perform a ``gauge transformation'' \cite{KA} to a new field $\tilde{Q}$
\begin{equation}
\label{g1}
Q_{t,t'}(x) =
\exp\left(i k_{\alpha}(x,t)\gamma^{\alpha}\right)
 \tilde Q_{t,t'}(x)
\exp\left(-i k_{\alpha}(x,t')\gamma^{\alpha}\right)\, .
                                                              \label{p9a}
\end{equation}
Introducing the long derivative
\begin{equation}
{\bf \partial}_x\tilde Q \equiv
\nabla \tilde Q +i [\nabla k_{\alpha}\gamma^{\alpha}, \tilde Q]\, ,
                                                            \label{q9}
\end{equation}

one may write the gradient expansion of   eq.~(\ref{act3}) as
\begin{eqnarray}
&&iS[\tilde Q,\Phi]=
i\nu \Tr \{(\Phi-i\omega K)^T \gamma_2 (\Phi+i\omega K) \} -
                                                             \label{q8}\\
&&\frac{\pi\nu}{4} \left[
D  \Tr \{ {\bf \partial}_x\tilde Q\}^2 +
4i \Tr \{(\epsilon +(\phi_\alpha+i\omega
k_{\alpha})\gamma^{\alpha}) \tilde Q \}
\right] \, \, .
                                                                \nonumber
\end{eqnarray}
At this point the  vector $K^T=(k_1,  k_2)$ that determines the
transformation (\ref{p9a}) is arbitrary. The saddle point equation
for $Q$  of the action (\ref{q8})
 is given by the  following equation
\begin{eqnarray}
\label{sadle_point2}
D \partial_x ( \tilde{Q} \partial_x\tilde{Q} ) +
i [(\epsilon +(\phi_\alpha+i\omega k_{\alpha})\gamma^{\alpha}),
\tilde{Q} ]  = 0 \, .
\end{eqnarray}
Let us now choose the parameterization
\begin{equation}
\label{par2}
\tilde{Q}=\tilde{\Lambda}\exp(\tilde{W}),
\end{equation}
where $\tilde{W}$ represents fluctuation
around the saddle-point
\begin{eqnarray}&&
\label{ansatz}
\tilde{\Lambda}(x,\epsilon)=\left(\matrix{1 & 2\tilde{F}[\phi](x,\epsilon)
\cr 0 & -1\cr}\right)\,\, .
\end{eqnarray}
Eq. (\ref{ansatz}) implies that the solution of the  saddle point equation 
(\ref{sadle_point2}), 
determines $\tilde{F}$ as a functional of $\phi$.
We do not know, though,  how to solve it.
Instead we average  over $\phi$ the eq.(\ref{sadle_point2}).
The solution of this averaged equation,  denoted by $\bar{F}$,  
is determined by:
\begin{equation}
\label{c3}
-D\nabla^2 \bar{F}(\epsilon)=I^{ee}\{F\}\, ,
\end{equation}
where the r.h.s. is given by
\begin{eqnarray}&&
\label{e29}
I^{ee}\{\bar{F}\}=D \int \frac{d\omega}{\pi}\big[
\langle \nabla k^1(\omega)\nabla\!k^1(-\omega)\rangle
\big(\bar{F}(\epsilon)-\bar{F}(\epsilon-\omega)\big)+
\nonumber \\&& 
\!(\langle\nabla\!k^1(\omega)\nabla\!k^2(-\omega)\rangle\!-\!\langle\nabla\!k^2(\omega)\nabla\!k^1(-\omega)\!\rangle)
\big(\bar{F}(\epsilon)\bar{F}(\epsilon\!-\omega)\!-\!1\big)
\big].
\end{eqnarray} 
Since ($\bar{\Lambda}$) is not a genuine saddle point of the 
action there is coupling  between the fields
$\tilde{W}$ and $\phi,(\nabla k)$ in the quadratic 
part of the action.
However the coupling constant between those fields 
is proportional to the  gradient of the distribution 
(cf. eq.(\ref{action_out_of_equlibrium})).
Therefore this term can be treated as a small perturbation.

Taking variation of the action with respect to $w$, $\bar{w}$, 
we obtain the following gauge, determining $k[\phi]$:
\begin{eqnarray}&&
\label{gauge}
D\nabla^2 k_2-\phi_2-i\omega k_2=0 \nonumber \\&&
D\nabla^2k_1+\phi_1+i\omega k_1=2\B[\omega,x]\nabla^2 k_2 \, ,
\end{eqnarray}
where  
\begin{eqnarray}&&
\B[\omega,x]=\frac{1}{2\omega}\int d\epsilon[1-\bar{F}(\epsilon,x)\bar{F}(\epsilon-\omega,x)]\, .
\end{eqnarray}
Though we have failed to find the true saddle point  the
linear part of the action expanded around ($\bar{\Lambda}$) is zero. 
It is remarkable to notice that under conditions (\ref{gauge})  
eq.(\ref{c3}) becomes a quantum kinetic equation \cite{quantum_kinetic_equation} with the  collision integral being $I^{ee}\{F\}$. 
Coming back to our calculations we note that the correlation function of
current fluctuations is a gauged invariant quantity (does not
depend on the position of the Fermi level). This means that
momenta $q \le \sqrt{\omega/D}$ do not contribute to   such a
quantity \cite{Fin}. In this case the Coulomb propagator is
universal, i.e. does not depend on the electron charge. The fact
that we address gauge invariant quantities  allows us to represent
the generating functional $Z$ in terms of the fields $Q$ and
$\nabla k$ (rather than $Q$ and $\phi$), as in Ref \cite{KA}.
\end{multicols}
\begin{eqnarray}&&
\label{c22}
\langle Z \rangle\!= \!
\int\!\! {\cal D}\nabla K\,
\exp\left( -i\nu D \Tr\{ \nabla K^T {\cal D}^{-1} \nabla K\}\right)
\int {\cal D} \tilde Q\,
\exp\left( \sum\limits_{l=0}^{2} i S_{l}[\tilde Q,\nabla K]\right) \, .
                                                              \label{u21}
\end{eqnarray}
Here we define
\begin{equation}
{\cal D}^{-1} =
\left(
\begin{array}{cc}
0 & -D\nabla^2_x  + i\omega \delta_{x,x'}  \\
-D\nabla^2_x  -  i\omega \delta_{x,x'} &
-2i\omega \delta_{x,x'}B_\omega(x)
\end{array}
\right)  \, \, ,
                                                              \label{k15}
\end{equation}
where  the  expansion $S=S^0+S^1+S^2$, is in powers of  $\nabla K$;
the $l-th$ power  ($ l=0,1,2$)  is
 given by

\begin{eqnarray}
\hspace{-1cm}iS^0[\tilde Q]=
-\frac{\pi\nu}{4}
\left[ D  \Tr \{ \nabla \tilde Q\}^2 +  4i\,  \Tr \{\epsilon \tilde  Q
\} \right],
\label{q11}
\end{eqnarray}                                                            
\begin{eqnarray}&&
\hspace{-1.2cm}
iS^1[\tilde Q,\nabla K]=-i\pi\nu\left[D\Tr\{\nabla k_\alpha\gamma^{\alpha} 
\tilde Q\nabla\tilde Q\}\!+\!\Tr\{(\phi_\alpha+i\omega k_\alpha)\gamma^{\alpha}\tilde Q\}\right],
\label{q12}
\end{eqnarray}    
\begin{eqnarray}&&
\hspace{-1.2cm}iS^2[\tilde Q,\nabla K]\!=\!
\frac{\pi\nu D}{2}\!
\left[\Tr\{\nabla k_\alpha\gamma^{\alpha}\tilde Q\nabla k_\beta\gamma^{\beta}\tilde Q\}\!-\!\Tr\{\nabla k_\alpha\gamma^{\alpha}\tilde{\Lambda}\nabla k_\beta\gamma^{\beta}\tilde{\Lambda}\}
\right].
\label{q13}
\end{eqnarray}

From eq.(\ref{k15}) we obtain the  gauge field correlation
function
\begin{eqnarray}&&
\label{cor_nabla_k}
\langle \nabla k_\alpha(x,\omega) \nabla k_\beta(x',-\omega) \rangle =
\frac{i}{D}Y_{\alpha,\beta}(\omega,x,x'),
\end{eqnarray}  
where
\begin{eqnarray}&&
\label{c4}
\hspace{-2.3cm}Y(\omega,x,x')\!\!=\!\!\left[\begin{array}{cc}
\!\!\!-2i\pi\nu\omega\int\!dx_1\D[\!-\omega,x,x_1]\B[\omega,x_1]\D[\omega,x_1,x']\!&
\!\D[-\omega,x,x']\!\!\! \\ \!\D[\omega,x,x']\!&\!0\!\!\!
\end{array}\right],
\end{eqnarray}
 Using eqs.(\ref{cor_nabla_k},\ref{c4}) we rewrite eq.(\ref{e29}) for the
quasi-one-dimensional wire as:
\begin{eqnarray}&&
\label{e36}
D\nabla^2 \bar{F}(\epsilon)=I^{ee}\{\epsilon,x\}\, ,
{\rm where }  \\&&
\label{c7}
I^{ee}(\epsilon,x)=\frac{i\pi}{2}\int d\omega
\big[
-2i\omega\pi\nu D[x,x_1,-\omega]B[\omega,x_1]\D[x_1,x,\omega]\big(\bar{F}(\epsilon)-\bar{F}(\epsilon-\omega)\big)+\nonumber \\&&
(\D[x,x,\omega]-\D[x,x,-\omega])(1-\bar{F}(\epsilon)\bar{F}(\epsilon+\omega))
\big].
\end{eqnarray}
\begin{multicols}{2}
The total number of particles and the total energy of the electrons
are both preserved during electron-electron and  elastic electron-impurity
scattering.  The collision integral, eq.(\ref{c7}),  satisfies then
\begin{eqnarray}&&
\label{c8}
\int_{-\infty}^{\infty}I^{ee}(\epsilon,x)d\epsilon=0,
\end{eqnarray}
\begin{eqnarray}&&
\label{c9}
\int_{-\infty}^{\infty}\epsilon I^{ee}(\epsilon,x)d\epsilon=0.
\end{eqnarray}
We now consider the limit   $l_{\rm ee} \ll L$.
The solution of eq.(\ref{c3})
assumes then the  form of a quasi-equilibrium single-particle distribution function
\begin{eqnarray}&&
\label{c12}
\bar{F}(\epsilon,x)=\tanh\left({\frac{\epsilon -e\phi(x)}{2T(x)}}\right).
\end{eqnarray}
Here $\epsilon$ is the total energy of the electron, and $e\phi$
is the electrostatic  potential and $T(x)$ is the
effective local  temperature of the electron gas.
To find the electrostatic  potential $\phi$
we employ eq.~(\ref{c8}). To facilitate our
calculations we further  assume that conductance band is symmetric
about the Fermi energy and that the spectral density of
single-electron energy levels is constant.
Integration over the energy, eq.(\ref{c8}) yields\\
\begin{equation}
\label{electrostatic}
{\partial_x}^2 \phi(x)=0.
\end{equation}
Solving  eq.(\ref{electrostatic}) 
under the condition that the voltage difference at the edges of a constriction is  $V$, we find  
\begin{equation}
\label{c6}
e\phi(x)=eV\left(\frac{x}{L}-\frac{1}{2}\right) +\bar{\mu}.
\end{equation}
Multiplying  eq.~(\ref{e36}) by energy and
and employing eq.(\ref{c9}) we obtain an equation
\begin{equation}
\label{eq_for_temperature}
\partial_x^2\bigg(\frac{\pi^2}{6}(kT(x))^2+\frac{1}{2}(e \phi(x))^2\bigg)=0.
\end{equation}
 The boundary condition of eq.(\ref{eq_for_temperature}) is
determined by the temperature of the electrons in the reservoirs.
Combining eqs.((\ref{c6})~and~(\ref{eq_for_temperature})) we find the electron
temperature in two opposite limits:
\begin{eqnarray}&&
\label{temperature} 
T(x)=\left\{
\begin{array}{l}
\frac{\sqrt{3}eV}{\pi L}\sqrt{x(L-x)} \ \ eV \gg T\,\, ,
\\
\ \ T, \hspace{2.3cm}eV \ll T\,\,.
\end{array}
\right.
\end{eqnarray}
Eqs.((\ref{c12}), (\ref{c6}) and (\ref{temperature}) determine the function
$\bar{F}$ uniquely.
We now replace the  right-corner element
of the matrix $\tilde{\Lambda}$ 
(i.e. $\tilde{F}[\phi]$, cf. eq.(\ref{ansatz}))
by its average value $\bar{F}$.

To calculate ${\cal{S}}_3$ under conditions of strong electron-electron
scattering (eq.~(\ref{inelastic})) one needs to replace the
operators $\hat{\I}^D$ and $\hat{\M}$  in eq.~ (\ref{z28})
by their gauged values
\end{multicols}
\begin{eqnarray}&&
\hspace{-2cm}
\S_3(t_1-t_2,t_2-t_3)=\frac{e^3(\pi\hbar\nu D)^2}{8}
\bigg\langle\frac{1}{2}
\hat{\tilde{\M}}(x_1,t_1)\hat{\tilde{\I}}^D
(x_2,x_3,t_2,t_3)+
(x_1,t_1\leftrightarrow\!x_3,t_3)+\nonumber \\&&
\hspace{-2cm}(x_1,t_1\leftrightarrow\!x_2,t_2)+
\frac{\pi\hbar\nu
D}{8}\hat{\tilde{\M}}(x_1,t_1)\hat{\tilde{\M}}(x_2,t_2)\hat{\tilde{\M}}(x_3,t_3)\bigg\rangle_{\nabla k, \tilde{Q}} \,\,  ,
\label{s3_in}
\end{eqnarray}
\begin{multicols}{2}
where the averaging is taken over the entire action $S$ and 
the Gaussian weight function for $\nabla K$, as in eq.~(\ref{gen2}).
Here we define (cf. eqs.(\ref{e33}), (\ref{M}) with eqs. (\ref{e332}),(\ref{c21}))
\begin{eqnarray}&&
\label{e332}
\hat{\tilde{\I}}^D(x,x',t,t')=\Tr\!\bigg\{\!\tilde{Q}_{x,t,t'}\gamma_2\tilde{Q}_{x',t',t}\gamma_2-\delta_{t,t'}\gamma_1\bigg\}\delta_{x,x'}
\,\, ,
\end{eqnarray}
\begin{eqnarray}&&
\label{c21}
\hat{\tilde{\M}}(x,t)=\Tr\!\bigg\{\!\int\!dt_1\bigg([\tilde{Q}_{x,t,t_1};\partial_x
]~ \tilde{Q}_{x,t_1,t} \bigg)\gamma_2\bigg\}\,\, ,
\end{eqnarray}
where the ``long derivative'', $\partial_x$, is presented in
eq.(\ref{q9}).
In order to actually perform the functional integration 
over the matrix field $\tilde{Q}$ we use
the parameterization of eq.~(\ref{par2}). 
We need  to find the Gaussian fluctuations around
the saddle point of the action (\ref{q11},\ref{q12},\ref{q13}).
Though we did not find the exact saddle point, the expansion of  $Q$ around 
$\bar{\Lambda}$ works satisfactorily. 
The coupling between the fields $\nabla k$ and $W$ which  
appears already in the Gaussian (quadratic) part is small, 
since it is proportional to the gradient of the distribution function:
\begin{eqnarray}&&
\label{action_out_of_equlibrium}
\!iS^1_1\!=\!-2i\pi\!g
\Tr\bigg\{\!
\bar{w}_{x,\epsilon,\epsilon'}
[\nabla k_{1 x,\epsilon'-\epsilon}
\nabla\bar{F}_{x,\epsilon}-
\nabla\bar{F}_{x,\epsilon'}\nabla\!k_{1 x,\epsilon'-\epsilon}\!+ \nonumber \\&&
\nabla\bar{F}_{x,\epsilon'}\nabla\!k_{2 x,\epsilon'\epsilon}\bar{F}_{x,\epsilon}+ \bar{F}_{x,\epsilon'}\nabla\!k_{2 x,\epsilon'-\epsilon}\nabla\bar{F}_{x,\epsilon}]
\bigg\}.
\end{eqnarray}
Here the upper index refers to the power of the  $\nabla k$
fields; the lower refers to the power  of $w, \bar{w}$ fields in
the expansion. 
Considered as a small perturbation, $iS_1^1$ does not affect the results.

The more dramatic effect on the correlation function arises from
the non-Gaussian  part of the action, eqs. (\ref{q12},\ref{q13})
(by this we mean non-Gaussian terms in either $w, \bar{w}$ or $\nabla K$).
After integrating over the interaction an additional contribution 
to the Gaussian part (proportional to $w\bar{w}$) of the action arises.
To find the effective action $iS^{\rm eff}[W]$ 
we  average over the interaction \cite{GG2}. 
One notes that 
\begin{eqnarray}&&
\label{identity1}
iS_0^1=iS_0^2=0\, ,
\end{eqnarray}
(where, again, $S_0^1$ refers to the component of the action, 
eq.(\ref{q12}), that has  zero power of the $w,\bar{w}$  fields and one power of the
$\nabla k$ field). 
In addition, due to the choice of the gauge, eq.(\ref{gauge}), and
the condition  ($l_{\rm ee} \ll L$), the averaging over $\nabla k$ does not 
generate  terms linear in $w,\bar{w}$ in the effective action:
\begin{eqnarray}&&
\label{identity2}
\langle iS^2_1\rangle_{\nabla k}=-2i\pi\nu\int\frac{d\epsilon}{2\pi}\bar{w}_{\epsilon,\epsilon} I_{\rm ee}[F]=0\, .
\end{eqnarray}
Combining eqs. (\ref{identity1} and \ref{identity2})
we find that the effective action acquires an additional contribution:  
\begin{eqnarray}&&
\bigg\langle \exp\left(iS^1+iS^2\right)\bigg\rangle_{\nabla k} \simeq 
\exp\left(\langle iS^2_2\rangle+\frac{1}{2}\langle iS^2_1 iS^2_1\rangle\right)
\end{eqnarray}
The general form of the effective action is rather complicated, however
for the low frequency noise only diagonal part of the action matters:
\begin{eqnarray}&&
\label{e55}
iS^{\rm eff}_2[w,\bar{w}]=
\frac{\pi\nu}{2}
\Tr\bigg\{\!\bar{w}_{x,\epsilon,\epsilon}
\!\bigg[\!-D\nabla^2\!+\!\hat{{\cal{I}}}^{ee}\bigg]w_{x,\epsilon,\epsilon}-
\nonumber \\&&
\bar{w}_{x,\epsilon,\epsilon}D\nabla\bar{F}_{x,\epsilon}\nabla\bar{F}_{x,\epsilon}\bar{w}_{x,\epsilon,\epsilon}
\!\bigg\}.
\end{eqnarray}
Here the operator
\begin{eqnarray}&&
\label{linearized_col_int}
\hat{{\cal{I}}}^{ee}w_{x,\epsilon,\epsilon}\equiv
\int d \omega[Y_{11}(\omega)[w_{\epsilon,\epsilon}-w_{\epsilon-\omega,\epsilon-\omega}]+\\&&
\left(Y_{12}(\omega)-Y_{21}(\omega)\right)
[F_\epsilon w_{\epsilon-\omega,\epsilon-\omega}+F_{\epsilon-\omega}w_{\epsilon,\epsilon}]+\nonumber \\&&
\int d\bar{\epsilon}\frac{1}{2\omega}
\left(F_{\epsilon}-F_{\epsilon-\omega}\right)
\left(Y_{12}(\omega)-Y_{21}(\omega)\right)
\left(F_{\bar{\epsilon}+\omega}+F_{\bar{\epsilon}-\omega}\right)
w_{\bar{\epsilon},\bar{\epsilon}} \nonumber 
\end{eqnarray}
is a {\it linearized} collision integral, 
i.e. a variation of the collision integral (\ref{c7}) 
with respect to the distribution function.
Substituting eqs.(\ref{e332}~and~\ref{c21}) into eq.(\ref{s3_in})
and calculating the Gaussian integrals with the action (\ref{e55}), we find 
\begin{eqnarray}&&
{\cal{S}}_3(\omega_1=0,\omega_2=0)=\frac{3e^3{\cal{A}} \pi g^2}{\hbar L^3}
\int_0^L d x_1 dx_2 \nonumber \\&&
\int_{-\infty}^{\infty}d\epsilon_1d\epsilon_2 \bar{F}(\epsilon_1,x_1)
{\cal{D}}[x_1,\epsilon_1;x_2,\epsilon_2]
\nabla\left(\bar{F}^2(\epsilon_2,x_2)\right) \, \, ,
\label{a10_inelastic}
\end{eqnarray}
where the ``inelastic diffusion'' propagator, ${\cal{D}}$,  is the
kernel the equation
\begin{eqnarray}&&
\label{inelastic_equation1}
\hspace{-0.5cm}[-D\nabla^2-\hat{{\cal{I}}}^{ee}]{\cal{D}}[x_1,\epsilon_1;x_2,\epsilon_2]\!=\!\frac{1}{\pi\nu}\delta(\epsilon_1-\epsilon_2)\delta(x_1-x_2).
\end{eqnarray}
For weak electron-electron scattering the collision integral
is small, yielding the standard propagator of the diffusion equation,
$\D$ (cf. eq. \ref{elastic_diffusion}).
In the presence of strong electron-electron scattering 
the collision integral dominates eq.(\ref{inelastic_equation1}).
In this limit we evaluate the leading asymptotic behavior 
for ($l_{\rm in}/L \ll 1$).
On scales longer than the inelastic mean free path 
the distribution function has a quasi-equilibrium form 
\begin{eqnarray}&&
f(\epsilon,x)=f_0\left( \frac{\epsilon-\mu(x)-\delta\mu(x)}
{T(x)+\delta T(x)}\right) ,
\end{eqnarray}
where
\begin{eqnarray}&&
f_0(x)=\frac{1}{1+\exp(x)}.
\end{eqnarray}
The  values of the local temperature and electro-chemical potential
can fluctuate. To find the  correlations of these fluctuations 
we consider the equation 
(with the same Kernel as in eq.(\ref{inelastic_equation1}))
\begin{eqnarray}&&
\label{inelastic_equation2}
[-D\nabla^2-\hat{{\cal{I}}}^{ee}]\delta f(\epsilon,x)=
\delta J(\epsilon,x) .
\end{eqnarray}
Integrating equation (\ref{inelastic_equation2}) with respect to energy
and using the particle-conservation property of the collision integral 
(eqs.(\ref{c8})) we find:
\begin{eqnarray}&&
\label{eq_on_delta_mu}
\mu(x)=\int dx'\D[x,x']\int_{-\infty}^{\infty}d\epsilon J(\epsilon,x') .
\end{eqnarray}
Using energy conservation (eq. \ref{c9})
we find
\begin{eqnarray}&&
\label{eq_on_delta_T}
\hspace{-0,5cm}\delta T(x)\!=\!\frac{3}{\pi^2}\frac{1}{T(x)}\int dx'\D[x,x']\int d\epsilon(\epsilon-\mu(x))\delta J(\epsilon,x').
\end{eqnarray}
Combining eqs. (\ref{eq_on_delta_mu}) and (\ref{eq_on_delta_T}) we find:
\begin{eqnarray}&&
\label{inelasticd}
{\cal{D}}[x_1,\epsilon_1;x_2,\epsilon_2]\!=\!\left(\!\frac{\partial}{\partial\epsilon}f_0\left(\frac{\epsilon_1\!-\!\mu(x_1)}{T(x_1)}\!\right)\!\right)\nonumber \bullet \\&&
\D[x_1\!,\!x_2]\!\left(\!-\!1\!-\!
\frac{3}{\pi^2}\frac{\epsilon_1\!-\!\mu(x_1)}{T(x_1)}
\!\frac{\epsilon_2\!-\!\mu(x_1)}{T(x_1)}\!\right) .
\end{eqnarray}

Evaluating ${\cal{S}}_3$ explicitly we find that the third
order current cumulant is
\begin{eqnarray}&&
{\cal{S}}_3(\omega_1=0,\omega_2=0)=\frac{36e^3Ag^2
eV}{L^4\pi}\int_0^L dx_1dx_2 \nonumber \\&&
\D[x_1,x_2]\bigg[\frac{T(x_1)}{T(x_2)}+(x_1-x_2)\frac{1}{T(x_2)}\frac{\partial}{\partial
x_1} T(x_1)\bigg].
\end{eqnarray}
At high temperatures (cf. eq.\ref{temperature}) one obtains
\begin{eqnarray}&&
{\cal{S}}_3(\omega_1=0,\omega_2=0)=\frac{3}{\pi^2}e^2I \, ,
\end{eqnarray}
while at low temperatures
\begin{eqnarray}&&
{\cal{S}}_3(\omega_1=0,\omega_2=0)=\left(\frac{8}{\pi^2}-\frac{9}{16}\right)
e^2I .
\end{eqnarray}
Our analysis was performed  for a simple rectangular
constriction. However, our results hold for any shape of the constriction, 
provided it is quasi-one dimensional
(we have considered a single transversal mode only).

\section{COMPARISON  WITH THE KINETIC THEORY OF FLUCTUATIONS}
\label{comparison}
\begin{figure}[h]
\includegraphics[width=0.7\textwidth]{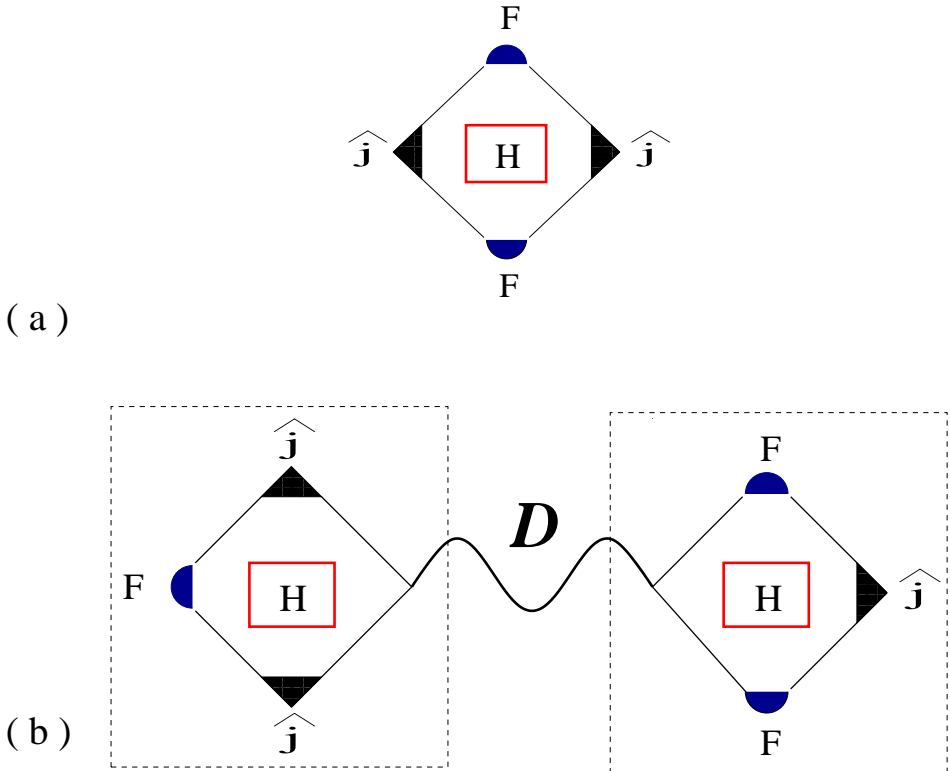}
\caption{The second (a) and the third (b) 
current correlation functions. 
\newline 
The vector vertices $\hat{j}$ represent the current operators, while 
collisions with static disorder correspond to the Hikami box ($H$).
$F=1-2f$, where $f$ is the single-electron distribution function.
D is the diffuson.}
\label{fig3}
\end{figure}

As was mentioned in Section  \ref{background} the applications of the kinetic equation
in the study of high order cumulants is not straight forward.
To compare our microscopic quantum mechanical 
calculation with the semi-classical 
Boltzmann-Langevin scheme we consider the diagrams
for the pair and third order correlation function, 
depicted in Fig.(\ref{fig3}).
   
The diagram in Fig. (\ref{fig3}-b) corresponds exactly to the results obtained above in the framework of the $\sigma$-model formalism.
Indeed, evaluating this diagrams, we recover the results  (\ref{a10}) and 
(\ref{a10_inelastic}).
Comparing Fig. (\ref{fig3}-b) with diagram (\ref{fig3}-a) 
(the latter determines the correlator  $\langle \delta J\delta J\rangle$ of Langevin sources), we conclude that the third 
cumulant corresponding to Fig. (\ref{fig3}-b) can be expressed in the form
\begin{eqnarray}&&
\label{conjecture}
\langle\delta\!J(1)\delta\!J(2)\delta\!J(3)\rangle= \nonumber \\&&
\hat{p}\int\!d4d4'
\frac{\delta \langle \delta J(1)\delta J(2)\rangle}{\delta f(4)}
D(4,4')\langle\delta\!J(4')\delta J(3)\rangle .
\end{eqnarray}

Indeed, the block on the right hand side of the diagram corresponds 
to the pair correlation function of random fluxes,
while the left part can be obtained by functional differentiation of the 
diagram  (\ref{fig3}-a) with respect to the function $F=1-2f$.
This justifies the  regression scheme proposed recently by Nagaev
\cite{Nagaev4}.
Diagram (\ref{fig3}-a) for the pair correlation function is
local in space; the diagram for the third order correlator is not.
Despite this non-locality the use of the  Langevin equation
(with non-local random flux) remains convenient; in that case 
the correlation function of the random fluxes 
needs to be calculated 
from first principles.
This is somewhat  analogous to the description of mesoscopic fluctuations
through the Langevin equation, 
proposed by Spivak and Zyuzin \cite{Spivak_Zuzin}.
\end{multicols}
\begin{multicols}{2}
Following Ref\cite{GG} the revisited version of the kinetic theory of fluctuations has been proposed by Nagaev \cite{Nagaev4}.
He has noticed that for a diffusive system
there exists a regression scheme of high order cumulants, expressed in terms
of pair correlators \cite{Nagaev5}.
However, a truly classical theory addressing  high moments of noise
needs to be taken  on the same footing as the Boltzmann-Langevin 
theory, i.e. without  appealing to  quantum mechanical
diagrammatics. Close inspection of the diagrams depicted 
in Fig. \ref{fig3} and the dimensionless  parameter by which corrections 
are small implies that the regression 
recipe (eq.~\ref{conjecture}) has the same range of applicability 
as the Boltzmann-Langevin scheme itself.

As a simple example of a classical problem
for which the regression procedure can be applied,  
we consider the following simple model\cite{Reznikov_private}.
A heavy molecule of mass $M$ and cross-section ${\cal{A}}$ 
is embedded in a gas of  light classical  particles. 
The gas consists of $N$ particles  of
mass $m \ll M$, in  thermal equilibrium with temperature $T$.
It is enclosed in a narrow tube of volume $V$.
The fluctuating velocity $u$ of the molecule here is the analogue  
of the fluctuating  electron distribution function $\delta f$.
The collisions of the molecule with the light particles is 
the counterpart of the electron's scattering on the random disorder.
The motion of the molecule is governed by Newton law
\begin{eqnarray}&&
\label{equation_of_motion}
M \frac{ d u}{d t}= F(t) .
\end{eqnarray}
Here $u$ is the velocity and $F$ is the force acting on the molecule.
The latter can be calculated from
\begin{eqnarray}&&
\label{force}
\!\!\!F(t)\!=\!2m{\cal{A}}\!
\bigg[\!\!\int_0^\infty\!\!\!dvv^2P_1(v\!+\!u(t),t)\!-\!\!\!\int_{-\infty}^0\!\!\!dvv^2P_2(\!v\!+\!u(t),t)\!\bigg],
\end{eqnarray}
where $P_1$  and $P_2$
are the fluctuating distribution functions of particles on
the left and on the right sides of the system respectively.
Since the time between two consecutive collisions is 
much shorter than the relaxation time of the molecule ($\tau_m$)
one can average over fast collisions while considering the relaxation 
dynamics
 of the heavy molecule:
\begin{eqnarray}&&  
\label{c2}
\frac{d \bar{u}}{dt}=-\frac{1}{\tau_m} \bar{u} .
\end{eqnarray}
To find the various  velocity correlation functions, one needs to know 
the  corresponding correlation functions of the random forces.
Using the fact that the equilibrium fluctuations of the distribution function 
in a Boltzmann gas are Poissonian (\cite{LL}, Section 114) and neglecting 
the small velocity $u$ relatively to $v$, 
one finds the  pair correlation function 
\begin{eqnarray}&&
\label{c88} 
\langle \delta F(t)\delta F(t')\rangle=
\frac{16}{\sqrt{2\pi}}\frac{AN}{V}m^2\langle v^2\rangle^{3/2}\delta(t-t') \, . 
\end{eqnarray}
Eq. (\ref{c88}) yields the correct value for the thermal fluctuations 
\mbox{$\langle u^2\rangle=T/M$.} To find  the higher order correlations
of the random force one {\it has} to take into account
the dependence of the  force on the velocity of the molecule.
We expand eq.(\ref{force}) up to second order in  $u$. 
After averaging over all possible pair correlations 
(such as $\langle u \delta P\rangle$ and $\langle \delta P \delta P\rangle $)
we obtain 
\begin{eqnarray}&& 
\label{c77}
\langle 
\delta F(t_1)
\delta F(t_2)
\delta F(t_3)
\rangle=\nonumber \\&&
\frac{512}{\sqrt{2\pi}}\left(\frac{AN}{V}\right)^2\!m^3
\!\langle u^2\rangle\langle v^2\rangle 
\hat{p}\!\bigg\{
\delta(t_1\!-\!t_2)\theta(t_2\!-\!t_3)\bigg\}.
\end{eqnarray}
Here $\hat{p}$ denotes  all  permutation over ($t_1,t_2,t_3$).
The triple force correlator was reduced to pair correlators only.
Bearing in mind the analogy between $u$ and $\delta f$ 
on one hand,  and the scattering of the electrons on the disorder and 
the scattering of the molecule by light particles on the other hand, 
we note that the reduction of the third order cumulant of the random forces
to the pair correlation functions  is similar in both cases.

We finally come back to the question of whether it is possible to calculate high order cumulants employing the classical kinetic equation
(Boltzmann-Langevin) rather than resorting to the diagrammatic
reduction scheme depicted above.
To be able to answer this question we first note that the applicability of 
the kinetic theory requires that  both 
$1/g, \hbar/\tau_{\rm corr}k_BT \ll 1$.
Here  $\tau_{\rm cor}$ is the correlation time for the current signal 
and $g$ is the dimensionless conductance.
For non-interacting electrons Onsager relation and 
Drude formula  yield 
$\tau_{\rm cor}=\tau$, where $\tau$ is the transport time.
It follows that the condition 
\begin{eqnarray}&&
\label{c18}
z\equiv \frac{\hbar}{g\tau_{\rm cor}k_BT} \ll 1
\end{eqnarray}
must be satisfied.
The reason for theses inequalities are first, that $g$ needs to be large in order for quantum interference effects to be negligible.
Secondly, the  value $\hbar/k_BT\tau$ needs also to be small; 
the  transport equation is applicable for times  longer than  
the duration time of an individual collision $\delta \tau$, 
($\tau \gg \delta \tau$).  
For a degenerate electron gas  the uncertainty relation 
requires that $\delta \tau \geq \hbar/k_BT$ \cite{Peierls,deviations}. 

We now evaluate the relative magnitude of, e.g.,
 the fourth and the second cumulants.
We consider the ratio 
\begin{eqnarray}&&
\tilde{z}\equiv \frac{<<I^4(0)>>}{<<I^2(0)>>^2} , 
\end{eqnarray}
where  $<< \ \ >>$ denotes the irreducible part of the correlator.
At equilibrium 
\begin{eqnarray}&&
\tilde{z} \simeq z .
\end{eqnarray}
As we see, for the kinetic theory to be valid,
the value of the parameter $z$ needs to be small.
But this is exactly the parameter ($\tilde{z}$) by which the high order 
correlation functions are smaller than the lower ones (cf. eq.\ref{c18}).
In other words, the evaluation of high order cumulants goes {\it beyond}
the validity of the standard Boltzmann-Langevin equation.
This is why we have to resort to the diagrammatic
approach: either to justify the reduction of high order cumulants to pair correlators, or, alternatively, to introduce non-local noise correlators 
in the Boltzmann-Langevin equation.
\section{COUNTING STATISTICS}
\label{counting}
So far we have studied the second and third  order cumulants.
In the present section  we discuss the whole distribution function
of the low frequency electron current (so-called counting statistics).
The zero temperature limit had been studied by Levitov et. al.
\cite{Levitov_Lee}. The full temperature regime was addressed 
by Nazarov\cite{Nazarov}.
Here we present a different derivation based on the $\sigma$-model 
approach.

Being a stochastic process,
the charge transmission can be characterized by the probability  
distribution function (PDF) $P_{\bar t}(n)$ of the  probability
for $n$ electrons to pass through the constriction 
within the time window $\bar t$.
In  practice it is more convenient to work with the Fourier 
transform of the PDF, the characteristic function  
\begin{eqnarray}&&
\kappa_{\bar t}[\lambda]=\sum_n e^{i\lambda n} P_{\bar t}(n) .
\end{eqnarray}
Below we evaluate $\kappa_{\bar t}[\lambda]$ for the case of elastic scattering.
By expanding the logarithm of characteristic function over its argument
we can find the cumulants (irreducible correlation functions) 
of a transmitted charge
\begin{eqnarray}&&
\label{generating_function1}
\ln(\kappa_{\bar{t}}[\lambda])=\sum_k\frac{(i \lambda)^k}{k!} {\cal{S}}_k .
\end{eqnarray}
For the problem of diffusive junction  the disorder average 
characteristic function can be represented as:  
\begin{eqnarray}&&
\label{charachteristic_function}
\bar{\kappa}_{\bar t}[\lambda]=\int {\cal{D}}Q \exp(iS[Q,{\bf a}]),
\end{eqnarray}
where the external source is given by 
\begin{eqnarray}&&
\label{external_source}
{\bf a}_2(t)=\left\{
\begin{array}{l}
\frac{\lambda}{2} \,\,\, ,\,\,\, 0<t<\bar t \\
0\,\,\,  , \,\,\,{\rm otherwise}
   \ \ .
\end{array} \right.
\end{eqnarray}
The applied bias enters the problem through  boundary conditions on $Q$ at the edges of the constriction:
\begin{eqnarray}&&
\Lambda(0,\epsilon)=\left(\matrix{1 & 2F\left(\epsilon+\frac{eV}{2}\right)
\cr 0 & -1\cr}\right) ,
\Lambda(L,\epsilon)=\left(\matrix{1 & 2F\left(\epsilon-\frac{eV}{2}\right)
\cr 0 & -1\cr}\right), \nonumber
\end{eqnarray}
and $F$ is defined by eq.(\ref{e31}).

Inasmuch as we are not interested in spatial correlations, 
the external source term is a function of time only.
Therefore, by performing the transformation
\begin{eqnarray}&&
Q(x,t,t')=e^{-ixa_2(t)\gamma_2}\tilde Q(x,t,t')e^{ixa_2(t')\gamma_2}.
\end{eqnarray}
one gets: 
\begin{eqnarray}&&
\label{action_gauged}
iS[\tilde{Q},{\bf a}]= -\frac{\pi\nu}{4\tau }\Tr\bigg\{D\left(\nabla\tilde{Q}\right)^2 -\left(\frac{\partial}{\partial t}+\frac{\partial}{\partial t'}\right)
\tilde{Q}(t,t')+
\nonumber \\&&
i\left(\frac{\partial}{\partial t}a_2(t)\right)\gamma_2
\tilde{Q}(t,t')-\tilde{Q}(t,t')i\frac{\partial}{\partial t'}a_2(t')\gamma_2\bigg\}\,\,,
\end{eqnarray}
and the boundary conditions change
correspondingly:
\begin{eqnarray}&&
\label{boundary_conditions2}
\tilde{Q}(0,t,t')=\Lambda(0,t-t'), \nonumber \\&&
\tilde{Q}(L,t,t')=e^{-ia_2(t)\gamma_2}\Lambda(L,t-t')e^{ia_2(t')\gamma_2},
\end{eqnarray}
with
\begin{eqnarray}&&
\Lambda(t-t')=\int \frac{d\epsilon}{2\pi}e^{i\epsilon(t-t')}\Lambda(\epsilon)\, .
\end{eqnarray}
We use the saddle point approximation to calculate the characteristic function,
Eq.~(\ref{charachteristic_function}).
In the presence of an  external potential the minimum of the  
action, eq.~(\ref{action_gauged}), satisfies
\begin{eqnarray}&&
\label{saddle_point_gauged}
D\nabla(\tilde{Q}\nabla \tilde{Q})-
\left(
\frac{\partial}{\partial t}+
\frac{\partial}{\partial t'}\right)\tilde{Q}+\nonumber \\&&
i\frac{\partial}{\partial t}a_2(t)\gamma_2\tilde{Q}(t,t')-\tilde{Q}(t,t')i\frac{\partial}{\partial t'}a_2(t')\gamma_2=0.
\end{eqnarray}

Let us define a parameter that can roughly be regarded  as ``the number of attempts per channel'', $M=\max\{eV,T\}{\bar t}/\hbar$.
We will focus on the case where the time window is much larger than the Thouless
time $\bar t \gg t_{Thouless}$ and the  conductance $g \gg M \gg 1$.
Inside the region $ 0<t,t'<\bar t $ eq.~(\ref{saddle_point_gauged})
becomes (up to the corrections $O(t_{\rm Thouless}/\bar{t}))$
\begin{eqnarray}&&
\label{simplified_sp_equation}
D\nabla(\tilde{Q}\nabla \tilde{Q})=0.
\end{eqnarray}
This can be represented in the form resembling a current conservation law
\begin{eqnarray}&&
\label{current_conservation}
D\nabla J =0 , 
\end{eqnarray}
where the   current is defined as
\begin{eqnarray}&&
J \equiv \tilde{Q}\nabla \tilde{Q}.
\end{eqnarray}
The solution of Eq.~(\ref{current_conservation}), 
$\tilde{Q}_{sp}(x)$ can be written as
\begin{eqnarray}&&
\tilde{Q}_{sp}=\tilde{Q}(0)\exp(Jx),
\end{eqnarray}
and from the boundary conditions (eq.~\ref{boundary_conditions2}) it follows that
\begin{eqnarray}&&
J=\ln(\tilde{Q}(0)\tilde{Q}(L)).
\end{eqnarray}
One may show that the anticomutator of the matrix $Q$ with $J$ vanishes
\begin{eqnarray}&&
\big\{\tilde{Q}(0),J\big\}_+=0.
\end{eqnarray}
Based on this property, we can show that the following statements do hold:
\newline
(1) The square of the matrix $\tilde{Q}$ is still equal to unity
\begin{eqnarray}&&
\tilde{Q}_{sp}^2(x)=1 ,
\end{eqnarray}
(2) The determinant of the matrix $\tilde{Q}$ satisfies
\begin{eqnarray}&&
\Det[\tilde{Q}_{sp}(x)]=-1,
\end{eqnarray}
(3) The trace of the matrix $\tilde{Q}$ is equal to zero
\begin{eqnarray}&&
\Tr\{\tilde{Q}_{sp}(x)\}=0,
\end{eqnarray}
(4) The trace of the matrix $J$ is zero
\begin{eqnarray}&&
\Tr\{J\}=0,
\end{eqnarray}
(5) The square of the matrix $J$ is proportional to the unit matrix
\begin{eqnarray}&&
J^2=\delta[\epsilon,\lambda]I,
\end{eqnarray}
where $\delta[\epsilon,\lambda]$ is given by (there is small discrepancy
with the result obtained in Ref.\cite{Nazarov})
\end{multicols}
\begin{eqnarray}&&
\delta[\epsilon,\lambda]=\ln\bigg[
 (2f_1-1)(2f_2-1)+2f_1(1-f_2)e^{-i\lambda}+2f_2(1-f_1)e^{i\lambda}+\nonumber \\&& 
 2\sqrt{
(e^{i\lambda/2}-e^{-i\lambda/2}) (1-f_1+f_1e^{-i\lambda}) (1-f_2+f_2e^{i\lambda}) (f_2(1-f_1) e^{i\lambda/2}-f_1(1-f_2) e^{-i\lambda/2})} \bigg] . \nonumber \\
\end{eqnarray}
\begin{multicols}{2}
Using these properties we find the disorder averaged counting statistics 
\begin{eqnarray}&&
\label{generating_function2}
\bar \kappa[\lambda]=\exp\left(
\frac{G{\bar t}}{\hbar}\int d\epsilon \delta^2[\epsilon,\lambda]
\right)
  .
\end{eqnarray}
The zero temperature limit 
coincides with the results derived previously by Lee et. al.\cite{Levitov_Lee}.

At finite temperatures we find:
\begin{eqnarray}&&
\label{zero_bias}
{\cal{S}}_2=2 G T \nonumber \\&&
{\cal{S}}_3=\frac{e^2}{3} I \nonumber \\&&
{\cal{S}}_4=\frac{2}{3} \ e^2 G T .
\end{eqnarray}
It is worthwhile to note that the value of ${\cal{S}}_4$ agrees
with the one obtained in Ref.\cite{Nagaev4}.
As we see the counting statistics of the current in a disordered wire 
is not Gaussian. 
Remarkably, {\it all the information} contained in the counting statistics
can be extracted from the pair correlation function 
(of the distribution function), eq.(\ref{int_eq6}). 

Finally we would like to discuss the role of 
{\it inelastic electron-phonon scattering}.
As has been already realized \cite{Blanter},
such an interaction suppresses shot noise.
Moreover, based on our approach,  
one can show that in the limit of $l_{\rm e-ph} \ll L$ (macroscopic conductor)
the current fluctuations are Gaussian 
(to leading order in $L/l_{\rm e-ph}$). 
To show this, we repeat our analysis concerning  ${\cal{S}}_3$.
We find that eq. (\ref{a10_inelastic}) still holds, 
but the distribution function and the inelastic diffusion propagator 
need to be calculated  taking  electron-phonon collisions into account.
In the presence of both electron-electron and electron-phonon interaction 
one obtains 
\begin{equation}
\label{c3_modified}
[D\nabla^2 \bar{F}+I^{ee}+I^{e-ph}]\bar{F}=0 \, ,
\end{equation}
(cf.~eq.\ref{c3}) where 
$\hat{{\cal{I}}}^{e-ph}$ is the linearized electron-phonon collision integral.
The inelastic diffuson (cf.~eq.\ref{inelastic_equation1}) is now determined by  
\begin{eqnarray}&&
\label{inelastic_equation3}
[\!-D\nabla^2-\hat{{\cal{I}}}^{ee}-\hat{{\cal{I}}}^{e-ph}]
{\cal{D}}[x_1,\epsilon_1;x_2,\epsilon_2]=\nonumber \\&&
\frac{1}{\pi\nu}\delta(\epsilon_1-\epsilon_2)\delta(x_1-x_2) .
\end{eqnarray}
In the limit $L/l_{\rm e-ph} \gg 1$
fluctuations of the chemical potential is the only long-range  
propagating mode in the problem 
(no fluctuations of the $k_BT$ along the system).
Solving eqs.(\ref{c3_modified}, \ref{inelastic_equation3} and 
\ref{a10_inelastic})
we find that ${\cal{S}}_3$ vanishes as $(l_{e-ph}/L)^2$.  
A conductor longer than the electron-phonon 
length ($l_{\rm e-ph}$) can be viewed as
a number ($L/l_{\rm e-ph}$)  of resistors connected in series. 
The current fluctuations of a single resistor are given by 
eq.(\ref{generating_function2}) and would render 
the corresponding voltage fluctuations.
Since the fluctuations at different resistors are {\it uncorrelated}
(local fluctuations in the chemical potential do not change the resistance
significantly)  the large number of the resistors results in 
{\it Gaussian}
 current fluctuations.
By contrast, temperature fluctuations (in the case of solely electron-electron  interactions) have long distance correlations, and give rise to non-Gaussian fluctuations.

\subsection*{Acknowledgments}
The authors wish to thank M. Reznikov and Y. Levinson  
for inspiring discussion and A. Kamenev for his critical 
comments in the earlier stages of this work.
This work was supported by GIF foundation,
the Minerva Foundation, the European RTN grant on spintronics,
by  the  SFB195 der Deutschen Forschungsgemeinschaft,
and by RFBR gr. 02-02-17688.

\end{multicols}

\begin{thebibliography}{99}
\bibitem{deviations}
As a side remark, we note that if the condition 
$\hbar/k_BT\tau \ll 1$ is not satisfied  the deviation 
from the kinetic theory may be substantial.
In particular the renormalization of conductance
by electron-electron interaction (Altshuler-Aronov effect) 
is pronounced when $\tau k_BT/\hbar$ is small\cite{AA}.
\end{thebibliography}
\end{document}